\documentclass[]{JHEP3}

\JHEPspecialurl{http://jhep.sissa.it/JOURNAL/JHEP3.tar.gz}

\usepackage{graphicx}
\usepackage{amsmath}
\usepackage{epsfig,multicol,bbm}

\newcommand{\be}{\begin{equation}}
\newcommand{\ee}{\end{equation}}
\newcommand{\bea}{\begin{eqnarray}}
\newcommand{\eea}{\end{eqnarray}}
\newcommand{\mbb}{\mathbb}

\newcommand{\mc}{\mathcal}

\newcommand{\beqa}{\begin{eqnarray}}
\newcommand{\eeqa}{\end{eqnarray}}

\newcommand\fverb{\setbox\fverbbox=\hbox\bgroup\verb}
\newcommand\fverbdo{\egroup\medskip\noindent%
			\fbox{\unhbox\fverbbox}\ }
\newcommand\fverbit{\egroup\item[\fbox{\unhbox\fverbbox}]}
\newbox\fverbbox

\title{Moduli Redefinitions and Moduli Stabilisation}

\author{Joseph P. Conlon, Francisco G. Pedro\\
	Rudolf Peierls Centre for Theoretical Physics, University of Oxford, 1 Keble Road, Oxford, OX1 3NP, UK\\
	E-mail: \email{j.conlon1@physics.ox.ac.uk},\email{f.pedro1@physics.ox.ac.uk}}

\preprint{OUTP-10-06P}	

\abstract{Field redefinitions occur in string compactifications at the one loop level. We review arguments for
why such redefinitions occur and study their effect on moduli stabilisation and supersymmetry breaking in the LARGE volume scenario.
For small moduli, although the effect of such redefinitions
can be larger than that of the $\alpha'$ corrections in both the K\"ahler and scalar potentials, they do not
alter the structure of the scalar potential. For the less well motivated case of large moduli, the redefinitions
can dominate all other terms in the scalar potential. We also study the effect of redefinitions on the structure of
supersymmetry breaking and soft terms.
}


\begin{document}

\section{Introduction}\label{sec:introduction}

In the study of string compactifications moduli stabilisation plays an essential role in attempts to generate a realistic
phenomenology. Moduli stabilisation is necessary to give masses to the light moduli that would otherwise give fifth forces, and is
also necessary to avoid runaway and decompactification. Furthermore the moduli potential plays a crucial role in setting both
the scale of supersymmetry breaking and the value of the gravitino mass, and so represents a necessary ingredient of any
ultraviolet-consistent model of supersymmetry breaking.

One set of promising models of moduli stabilisation are the LARGE volume models originally developed in \cite{hepth0502058, hepth0505076}.
The main attractive
feature of these is that they both stabilise all moduli and dynamically break supersymmetry at exponentially small scales.
In this way they address one of the main obstacles to stringy phenomenology and also one of the outstanding issues of BSM
physics, namely how to generate hierarchies.

An important role in the LARGE volume models is played by corrections to the K\"ahler potential \cite{BBHL}. The minimum of the scalar potential
comes from competition between a perturbative $\alpha'$ correction to the K\"ahler potential and non-perturbative corrections
to the superpotential. While the form of the superpotential is restricted by non-renormalisation theorems, the K\"ahler potential
is not constrained. To analyse the robustness of the LARGE volume scenario it is therefore important to check the effect of
all other possible corrections to the K\"ahler potential.

These corrections fall into several forms. The most obvious kind are those of other $\alpha'$ corrections. The LARGE volume models
involve the correction to the K\"ahler potential induced by the $\mc{R}^4$ term present at $\mc{O}(\alpha'^3)$ in the 10d IIB
action. There are many other additional terms (for example flux terms such as $G^2 \mc{R}^4$). Such $\alpha'$ terms were considered
in \cite{hepth0505076}. At large volume such terms are all subleading compared to the $\mc{R}^4$ term, essentially because the
$\mc{R}^4$ term is suppressed by six powers of length, while flux terms are suppressed by higher powers (as flux is quantised on
multi-dimensional cycles).

Another form of correction are those induced by string loop corrections to the K\"ahler potential. For some toroidal models
these have been calculated in \cite{0508043, 0508171} (also see \cite{0507131}).
For more general models the Coleman-Weinberg potential has been used to
constrain the form and magnitude of loop corrections \cite{07081873, 08051029, 09070665}.
One interesting property of such loop corrections is that they do in
general induce corrections in the K\"ahler potential that dominate at large volume over the $\alpha'^3$ corrections. However,
due to an extended no-scale structure such loop corrections in fact cancel at leading order in the scalar potential \cite{07081873}, remaining
sub-dominant to the $\alpha'^3$ corrections. Other studies of the robustness of the LARGE volume scenario include
\cite{07040737, 07070105, 08041248}.

This paper will study another form of possible correction that does not seem to fall neatly into either of the classes above.
This refers to the effect of field redefinitions that can occur at the loop level. In such redefinitions the relationship
between the holomorphic moduli and the physical cycle volumes is modified at 1-loop, such that the holomorphic moduli no longer
correspond directly to cycle sizes. As the K\"ahler potential involves the physical volume this should lead to a correction
to the K\"ahler potential as the expression for the physical volumes in terms of the moduli changes.

The existence of such redefinitions has been established for various orbifold models where exact string computations are possible.
It is certainly plausible that such redefinitions survive beyond the orbifold limit in which they were computed. We shall assume this
applies, and then study the effect of redefinitions on moduli stabilisation and the computation of soft terms. The structure of the
paper is as follows. In section \ref{sec2} we review the LARGE volume scenario and the particular model we study, defined by a hypersurface in $\mbb{P}^4_{[1,1,1,6,9]}$. In section 3 we review how field redefinitions occur in orbifolds and give a new argument for their
occurrence in the geometric regime. In section 4 we study the effect of such redefinitions on moduli stabilisation, and in section 5
we extend this study to the effects on soft supersymmetry breaking. In section 6 we conclude.

\section{The LARGE Volume Scenario}
\label{sec2}

We start by reviewing the basic features of the LARGE Volume Scenario for Type IIB orientifolds, proposed in \cite{hepth0502058}.
Neglecting gauge interactions,
 a 4-dimensional supergravity model is specified entirely by the K\"ahler potential and the superpotential.
 To leading order in $g_s$ and $\alpha'$ the K\"ahler potential is given by \cite{hepth0403067}
\begin{equation}
	\hat{K}=-2 \ln [\mathcal{V}]-\ln \left[-i \int_\mathcal{M} \Omega\wedge\bar{\Omega}\right]- \ln[-i(\tau-\bar{\tau})]
	\label{eq:KnoScale}
\end{equation}
where $\mathcal{V}$ is the volume of $\mathcal{M}$ measured in the Einstein frame ($g_E=e^{-\phi/2}g_s$) in units of the string length $l_s=2\pi\sqrt{\alpha^{'}}$. The volume is given by
\begin{equation}
	\mathcal{V}=\frac{1}{6}\int_\mathcal{M}J\wedge J \wedge J= \frac{1}{6} k_{ijk} t^i t^j t^k,
	\label{eq:}
\end{equation}
where $J$ is the K\"ahler form of $\mathcal{M}$, $k_{ijk}$ are triple intersection numbers and $t^i$ are two-cycle volumes. The fields appearing in the 4D effective theory are the complexified K\"ahler moduli $T_i\equiv - i \rho_i=\tau_i + i b_i$ where $b_i$ is the axion and the $\tau_i$ are the Einstein frame four-cycle volumes, given by:
\begin{equation}
	\tau_i=\frac{\partial \mathcal{V}}{\partial t^i}=\frac{1}{2}k_{ijk}t^j t^k.
	\label{eq:}
\end{equation}
The tree level superpotential is given by \cite{9906070, gkp}
\begin{equation}
		\hat{W}=\int_\mathcal{M} G_3\wedge\Omega,
	\label{eq:Wflux}
\end{equation}
where $\Omega$ is the (3,0) form of the Calabi-Yau and $G_3=F_3- \tau H_3$. It is important to note that $\hat{W}$ is independent of the K\"ahler moduli.

The class of models defined by Eqs. (\ref{eq:KnoScale}) and (\ref{eq:Wflux}) are examples of no-scale models. The $\mathcal{N}=1$  supergravity scalar potential is
\begin{equation}
	V=e^K (G^{i \bar{j}} D_i W D_{\bar{j}}\bar{W} -3 |W|^2).
	\label{eq:V}
\end{equation}
Computing the scalar potential using the K\"ahler potential and the superpotential of Eqs. (\ref{eq:KnoScale}) and (\ref{eq:Wflux})  one finds that Eq. (\ref{eq:V}) reduces to
\begin{equation}
	V_{no-scale}=e^K (G^{a \bar{b}} D_a W D_{\bar{b}}\bar{W} ),
	\label{eq:VnoScale}
\end{equation}
where the sum runs over only the dilaton and the complex structure moduli.
The  K\"ahler moduli dependence in Eq. (\ref{eq:VnoScale}) appears just as an overall factor.
$V_{no-scale}$ is positive definite and fixes the dilaton and complex structure moduli at a locus given by
\begin{equation}
	D_a \hat{W}\equiv \partial_a \hat{W}+\hat{W}\partial_a K=0.
	\label{eq:DaW}
\end{equation}
The value of the superpotential when (\ref{eq:DaW}) is satisfied is denoted by $W_0$.

The no-scale structure is a tree-level result and is broken by both $\alpha'$ and $g_s$ effects.
In \cite{BBHL} it was shown that the inclusion of $\mathcal{O}(\alpha'^3)$ terms in the K\"ahler potential leads to the  breaking of the no-scale structure of the scalar potential:
\begin{equation}
		\hat{K}=-2 \ln \left[\mathcal{V}+\frac{\xi (\frac{S+\bar{S}}{2})^{3/2}}{2 } \right]-\ln[-i \int_M \Omega\wedge\bar{\Omega}]-\ln[-i(\tau-\bar{\tau})],
	\label{eq:KBBHL}
\end{equation}
where $\xi=-\frac{\zeta(3)\chi(\mathcal{M})}{2 (2 \pi)^3}$ with $\chi(\mathcal{M})=h^{2,1}-h^{1,1}$. We focus on the case $\xi>0$ which is equivalent to having more complex structure than K\"ahler moduli.
The no-scale structure can also be broken by (non-perturbative) corrections to the superpotential \cite{kklt}, giving it the form
\begin{equation}
		\hat{W}=\int_\mathcal{M} G_3 \wedge \Omega + \sum_i A_i e^{i a_i \rho_i}.
	\label{eq:W}
\end{equation}

In \cite{hepth0502058} it was shown that the combination of these two effects
can generate minima of the scalar potential at exponentially large volumes.
To study these we first stabilise the dilaton and the complex-structure moduli by solving Eq. (\ref{eq:DaW})
and then regard their values as fixed (this approximation is justified \emph{post hoc} by the large value of
the stabilised volume, which gives minimal cross-coupling between these sectors).

 The resulting theory is only dependent on the K\"ahler moduli and is specified by
\begin{equation}
	\hat{K}=-2 \ln[\mathcal{V}+\frac{\xi}{2 g_s^{3/2}}]+K_{cs},
	\label{eq:KLVS}
\end{equation}
and
\begin{equation}
		\hat{W}=W_0+\sum_i A_i e^{- a_i T_i}.
	\label{eq:WLVS}
\end{equation}

 The original
 discussion of the minima of the potential in Eq. (\ref{eq:V}) for a generic compactification manifold can be found in \cite{hepth0502058},
 establishing the existence of an AdS minimum at exponentially large volumes.
 Subsequent discussions extending this to more general manifolds can be found in \cite{08051029}. For the purposes of this paper
 we will focus on the original model used in \cite{hepth0502058}, namely a degree 18 hypersurface in $ \mbb{P}^4_{[1,1,1,6,9]}$.
 For reference we briefly review this case.
 In terms of the 4-cycles the volume of the manifold is given by
\begin{equation}
	\mathcal{V}= \frac{1}{9\sqrt{2}}(\tau_b ^{3/2}-\tau_s^{3/2})\equiv \frac{1}{\lambda}(\tau_b ^{3/2}-\tau_s^{3/2}).
	\label{eq:Vol}
\end{equation}

Equations (\ref{eq:KLVS}), (\ref{eq:WLVS}) and (\ref{eq:Vol}) allow us to compute the scalar potential from Eq. (\ref{eq:V}) obtaining
\begin{equation}
	V= \frac{8}{3}\frac{\lambda a^2 |A|^2}{\mathcal{V}}e^{-2 a \tau_s}\sqrt{\tau_s}-4 \frac{|A W|}{\mathcal{V}^2}a\tau_s e^{-a \tau_s}+\frac{3}{4}\frac{|W|^2\xi}{\mathcal{V} ^3 g_s^{3/2}}.
	\label{eq:V2}
\end{equation}

We want to find the minimum of this potential and to study its properties. For that one must solve $\frac{\partial V}{\partial \tau_b}=\frac{\partial V}{\partial \tau_s}=0$. The first condition yields
\begin{equation}
	\frac{\partial V}{\partial \mathcal{V}}=0\Leftrightarrow \mathcal{V}=\frac{3 |W|}{2 \lambda a A}\sqrt{\tau_s}e^{a \tau_s}\left( 1\pm\sqrt{1-\frac{3 \lambda \xi}{8 g_s^{3/2}\tau_s^{3/2}}}\right),
	\label{eq:dvdv}
\end{equation}
while the second
\begin{equation}
	\frac{\partial V}{\partial \tau_s}=0\Leftrightarrow(\frac{1}{2}-2 a \tau_s)\left( 1\pm\sqrt{1-\frac{3 \lambda \xi}{8 g_s^{3/2} \tau_s^{3/2}}}\right)=(1-a \tau_s).
	\label{eq:dvdts}
\end{equation}
In the limit of large $a \tau_s$ Eq. (\ref{eq:dvdts}) becomes
\begin{equation}
	\tau_s^{3/2}\approx\frac{\lambda \xi}{g_s^{3/2}}\left(\frac{1}{2}+\frac{1}{4 a \tau_s}+\mathcal{O}(a \tau_s)^{-2}\right).
	\label{eq:taus}
\end{equation}
Equation (\ref{eq:taus}) gives an implicit solution for $\tau_{s}$, one can get an explicit expression by solving it iteratively. The first iteration will be
\begin{equation}
	\tau_s^{3/2}\approx\tilde{\tau}_s^{3/2}\left(1+\frac{1}{2 a \tilde{\tau}_s}\right),
	\label{eq:taus2}
\end{equation}
where $\tilde{\tau}_s^{3/2}=\frac{\lambda \xi}{2 g_s^{3/2}}$.
We will use eq. (\ref{eq:taus2}) as our analytical expression for $\tau_s$ at the minimum, neglecting the small
subleading $\mc{O}\left( a^{-2} \tau_s^{-2} \right)$ corrections.
We can also expand Eq. (\ref{eq:dvdv}) in powers of $\frac{1}{a \tau_s}$ to find:
\begin{equation}
	\mathcal{V}=\frac{3 |W_0|}{4 \lambda a |A|} \sqrt{\tau_s} e^{ a \tau_s}\left(1-\frac{3}{4 a \tau_s}+\mathcal{O}(a^{-2}\tau_s^{-2})\right).
	\label{eq:dvdv2}
\end{equation}



\section{1-loop K\"ahler Moduli Redefinition}\label{sec:kahler_moduli_redefinition}

The purpose of this paper is to study the effects of field redefinitions on moduli stabilisation. It is a known feature of
string compactifications that the definitions of moduli are altered at 1-loop level. Specifically, what is altered is the definition of the
chiral superfields of the supergravity action in terms of the geometric quantities of the compactification.

The LARGE volume models involve a large overall volume together with smaller blow-up cycles. If such blow-up cycles were collapsed
to a point, they would generate a singularity. Branes at singularities were recently studied in the context of threshold corrections
in \cite{09014350, 09061920}. It was found there that in various cases consistency with the effective supergravity - specifically
the Kaplunovsky-Louis formula describing the effect of anomalous contributions to gauge coupling running -
required that the modulus controlling the
blow-up cycle had to be redefined at 1-loop level. For other examples of 1-loop redefinitions in the context of theories
with D-branes, see \cite{9906039,07052366}. We emphasise that here we focus on the role of K\"ahler moduli in the redefinition; the dilaton
and complex structure moduli will also appear, but as these are flux-stabilised redefinitions will be less important.

The redefinition did not occur in all cases (for example it occurred for D3 branes at orientifolded singularities, but not at orbifolded singularities). Where it did occur the redefinition took the form of a shift by a
factor proportional to the logarithm of the overall volume of the CY:
\begin{equation}
		\tau_{new} = \tau_{old} - \alpha \ln (\mathcal{V}),
		\label{eq:redef}
\end{equation}
where $\alpha$ is taken to be small. The physical cycle volume (which vanishes at the singularity)
corresponds to $\tau_{old}$. The holomorphic modulus $\tau_{new}$ is however non-zero at the singularity.
We also note the types of cycles for which the redefinition occurs (blow-up cycles) are essentially the same as the
blow-up moduli that play a crucial role in the LARGE volume construction.

While such orbifold calculations show that field redefinitions occur, they are restricted to the singular limit.
Let us give another argument for why field redefinitions should take place, this time in the geometric regime. Consider an isolated collapsible cycle
(e.g. a $\mbb{P}^2$) taken to itself under an orientifold action, such that an O-plane wraps it
 and a stack of D-branes is on top
of the O-planes giving a rigid $\mc{N}=1$ SYM theory.
This is the type of dynamics required on the small cycle in the LARGE volume scenario.
Such a theory will undergo gaugino condensation and generate a non-perturbative
superpotential. The classical tree-level gauge coupling is given by
\be
\frac{4 \pi}{g^2} = \tau,
\ee
where $\tau$ is the Einstein frame
physical volume of the 4-cycle measured in units of the string
length ($e^{-\phi} \hbox{Vol}_{\Sigma}/(2 \pi \sqrt{\alpha'})^4$). The running gauge coupling is then given by
\be
\frac{1}{g^2}(\mu) = \frac{\tau}{4 \pi}+ \frac{\beta}{16 \pi^2} \ln \left( \frac{\Lambda_{UV}^2}{\mu^2} \right),
\ee
and so $\Lambda_{strong} = \Lambda_{UV} e^{-\frac{2 \pi \tau}{\beta}}$.
The condensing gauge group also generates a holomorphic superpotential
\be
\label{super1}
W = M_P^3 e^{-\frac{3}{\beta} (2 \pi T)}.
\ee
Here $T$ is the chiral superfield corresponding to the 4-cycle wrapped by the branes, given
classically by $T = \tau + i \int_{\Sigma} C_4$, but as we will
see this does not have to hold at loop level.
The prefactor of $M_P^3$ in (\ref{super1}) is enforced
by supergravity: $M_P$ is the only dimensionful quantity present in 4d supergravity and so is the only
quantity that can appear here. In particular, quantities such as the string scale $M_s = M_P/\sqrt{\mc{V}}$ cannot appear
in (\ref{super1}) as they violate holomorphy.

However, the normalised superpotential should also be identified with the strong coupling scale:
\be
e^{K/2} W = \left\langle \bar{\lambda} \lambda \right\rangle = \Lambda^3_{strong}.
\ee
Using $\mc{K} = - 2 \ln \mc{V}$, we therefore obtain
\be
\label{consistent}
\Lambda_{strong} = \Lambda_{UV} e^{-\frac{2 \pi \tau}{\beta}} = \frac{M_P}{\mc{V}^{1/3}} e^{-\frac{2 \pi \textrm{Re}(T)}{\beta}}.
\ee
In principal two possibilities can now occur.
First, the UV scale from which couplings start running could be identified with $M_P/\mc{V}^{1/3}$. Indeed it is known
\cite{09014350,09061920} that in various local models this is actually the scale from which couplings start running.
However the detailed analysis of \cite{09014350,09061920} also reveals that the physics which effectively allows couplings to start running at
a scale $M_W \equiv \frac{M_P}{\mc{V}^{1/3}} = R M_S$ corresponds to uncancelled RR tadpoles in the $\mc{N}=2$ sector.
These are tadpoles that are sourced locally but cancelled globally, and geometrically correspond to cycles
which have homology representatives both locally and also in the bulk.
However this does not apply here, where
the cycle is purely local (it involves purely $\mc{N}=1$ sectors; equivalently the cycle is dynamical in the local geometry).
This means we should identify $\Lambda_{UV}$ with $M_{string}$, in which case
consistency of equation (\ref{consistent}) requires that
\be
\hbox{Re}(T) = \tau + \frac{\beta}{2 \pi} \ln \mc{V}^{1/6}.
\ee
As $\tau$ involves a factor of $e^{-\phi}$ from the Einstein frame volume,
this represents a 1-loop redefinition of $T$ which appears to
be required for consistency with the framework of 4d supergravity.

Both the above arguments suggest
moduli redefinitions may be occurring in geometries relevant to moduli stabilisation. We therefore think it is important to study the possible
effects on moduli stabilisation. While the form of the
superpotential must be unaltered (as before we focus on the dependence on the K\"ahler moduli; the dependence on the dilaton and
complex structure moduli can change), the redefinitions can affect the K\"ahler potential. For the purpose of this paper
we shall make a simple assumption as to the nature of these affects. The classical K\"ahler potential is given by
$K = - 2 \ln \mc{V}$, where $\mc{V}$ is the physical (Einstein frame) volume of the Calabi-Yau. In this paper we shall assume that
the dependence on the physical volume remains the same, and that the change in the K\"ahler potential as a function of the moduli
comes simply from re-expressing the volume in terms of the redefined moduli. This assumption could be verified by a 1-loop computation of
the K\"ahler potential, which is however beyond the scope of this work.

We note both arguments relate to redefinitions of the small blow-up modulus. The motivation for redefinitions of the large
modulus is weaker. Indeed in this case the explicit loop calculation of \cite{0508043} does not show any evidence for it
(as this was a toroidal computation, it should incorporate the behaviour of the overall volume modulus).
For completeness we shall however also study redefinitions of this large modulus in the following section.

\section{Moduli Stabilisation}\label{sec:moduli_stabilisation}

In this section we investigate the consequences of 4-cycle volume redefinition in moduli stabilisation. We focus on the particular case of the $\mbb{P}^4_{[1,1,1,6,9]}$ orientifold and study what happens to the position of the minimum, Eqs. (\ref{eq:taus2}) and (\ref{eq:dvdv2}), when we redefine $\tau_b$ and $\tau_s$. We study these two cases separately.

\subsection{Redefining $\tau_s$}

Refedining $\tau_s$ according to Eq. (\ref{eq:redef}) we find that the K\"ahler potential for moduli fields becomes:
\begin{equation}
	K=-2 \ln\left[\frac{1}{\lambda}(\tau_b^{3/2}-[\tau_s-\alpha \ln(\mathcal{V})]^{3/2})+\frac{\xi}{2}\left(\frac{S+\bar{S}}{2}\right)^{3/2}\right] + K_{cs}.
	\label{eq:Kredeftaus}
\end{equation}
We allow an arbitrary parameter $\alpha$ in the redefinition. Note the term proportional to $\alpha$ dominates at large volume
over the term proportional to $\xi$.
Computing the supergravity scalar potential we find
\begin{equation}
	\begin{split}
	V&=\frac{1}{\tau_b^{9/2}}\left(-\frac{9}{2}|W_0|^2 \alpha \sqrt{\tau_s} \lambda^2 +\frac{3}{4}|W_0|^2 \frac{\lambda^3 \xi}{g_{s}^{3/2}}  \right)+\frac{2}{3}\frac{a^2 |A|^2 \lambda^2 e^{-2 a \tau_s}}{\tau_b^{3/2} \sqrt{\tau_s}}(4 \tau_s -2 \alpha \ln(\mathcal{V}))-\\
	&-\frac{1}{\tau_b ^3}a \left(\bar{A} e^{a \tau_s}W_0+ A e^{a \tau_s} \bar{W}_0\right)e^{-2 a \tau_s}\lambda^2 (2\tau_s+3 \alpha-2\alpha \ln(\mathcal{V})).
	\end{split}
	\label{eq:Vredeftaus}
\end{equation}
We have dropped terms that are subleading by powers of $\mc{V}$.
This can be further simplified to:
\begin{equation}
	\begin{split}
	V=&\frac{3}{4}|W_0|^2\frac{\xi \lambda^3}{g_{s}^{3/2}}\left(1-\frac{6\alpha \sqrt{\tau_s}}{\xi \lambda}\right)\frac{1}{\tau_b^{9/2}}+\frac{8}{3}\frac{a^2 |A|^2 \lambda^2 e^{-2 a \tau_s}}{\tau_b^{3/2}}\sqrt{\tau_s}\left(1- \frac{\alpha}{2\tau_s} \ln(\mathcal{V})\right)-\\
	&-\frac{\lambda^2}{\tau_b ^3} 2 a \tau_s \left(\bar{A}W_0+A\bar{W}_0\right)e^{- a \tau_s} \left(1-\frac{\alpha}{\tau_s} \ln(\mathcal{V})\right),
	\end{split}
	\label{eq:Vredeftaus}
\end{equation}
by noting that at the minimum we expect $\ln\mathcal{V}=a \tau_s \gg1$ and that the nonperturbative part of the superpotential is small compared to $W_0$. One can rewrite Eq. (\ref{eq:Vredeftaus}) as
\begin{equation}
	V=\Lambda\left(1-\frac{6\alpha g_s^{3/2}\sqrt{\tau_s}}{\xi \lambda}\right)\frac{1}{\tau_b^{9/2}}+\Omega \frac{\sqrt{\tau_s} e^{-2 a \tau_s}}{\tau_b^{3/2}}\left(1-\frac{\alpha}{2 \tau_s} \ln\mathcal{V}\right)-\Psi \frac{\tau_s e^{-a \tau_s}}{\tau_b^3}\left(1-\frac{\alpha}{\tau_s} \ln\mathcal{V}\right).
	\label{eq:}
\end{equation}
by defining
\begin{eqnarray}
	\label{eq:Lambda}
\Lambda & = & \frac{3}{4} |W_0|^2 \lambda^3 \xi/g_s^{3/2} ,\\
	\label{eq:Omega}
\Omega & = &  \frac{8}{3} a^2 |A|^2 \lambda^2,\\
	\label{eq:Psi}
\Psi & = &  2 a \lambda^2 (\bar{A}W_0+\bar{W_0}A).
\end{eqnarray}
At this point we again note that the contribution from the field redefinition is dominant over the $\alpha'^3$
correction also in the scalar potential as well as the K\"ahler potential. This makes the study of this potential
particularly interesting as the $\alpha'^3$ correction played an important role in establishing the existence of the
original LARGE volume minimum.

\begin{figure}[ht]
	\centering
	\begin{minipage}[b]{0.45\linewidth}
	\centering
	\includegraphics[width=\textwidth]{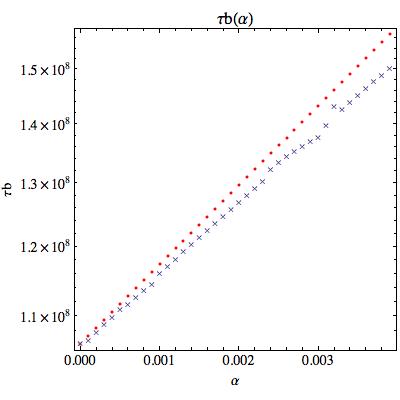}
	\label{fig:TAUbPlots}
	\end{minipage}
	\hspace{0.5cm}
	\begin{minipage}[b]{0.45\linewidth}
	\centering
	\includegraphics[width=\textwidth]{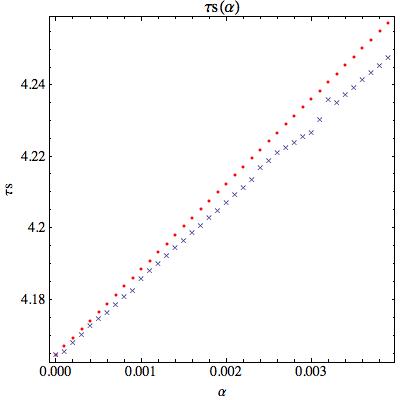}
	\label{fig:TAUsPlots}
	\end{minipage}
	\caption{$\tau_b$ and $\tau_s$ at the minimum as functions of $\alpha$. The dots are the analytical solutions while the crosses represent the numerical results.}
	\label{fig:TAUbsPlots}
	\end{figure}

To find the minimum of the potential one must solve $\frac{\partial V}{\partial \tau_s}=	\frac{\partial V}{\partial \tau_b}=0$.
Keeping in mind that we expect $\ln\mathcal{V}\approx a \tau_s \gg 1$, from the first condition we find that at the minimum
\begin{equation}
	\tau_b^{3/2}=\frac{\Psi}{2 \Omega}\sqrt{\tau_s}e^{a \tau_s}\left(1- \frac{3}{4 a \tau_s}-\frac{\alpha}{2 \tau_s}\ln\mathcal{V}\right),
\end{equation}
while the second condition yields:
\begin{equation}
	\tau_s=\left(\frac{4\Lambda\Omega}{\Psi^2}\right)^{2/3}\left(1+ \frac{1}{3 a \tau_s}+\alpha\left(\frac{\ln\mathcal{V}}{\tau_s}-\frac{4 g_s^{3/2} \sqrt{\tau_s}}{\lambda \xi}\right)\right).
\end{equation}
Using the definitions in Eqs. (\ref{eq:Lambda})-(\ref{eq:Psi}) we may rewrite these equations as:
\begin{eqnarray}
	\tau_s&=&\tilde{\tau}_s\left(1+ \frac{1}{3 a \tau_s}+\alpha\left(\frac{\ln\mathcal{V}}{\tau_s}-\frac{2}{\tilde{\tau}_s}\right)\right),\\
	\label{eq:}
	\tau_b&=&\tilde{\tau}_b\left(1- \frac{1}{2 a \tau_s}-\frac{\alpha}{3 \tau_s}\ln\mathcal{V}\right),
	\label{eq:}
\end{eqnarray}
where $\tilde{\tau}_s$ and $\tilde{\tau}_b$ are given by:
\begin{eqnarray}
	\label{eq:taustilde}
	\tilde{\tau}_s&=& \left(\frac{\lambda \xi}{2 g_s^{3/2}}\right)^{2/3},\\
	\tilde{\tau}_b&=&\left(\frac{3|W_0|}{4|A|a}\sqrt{\tau_s}e^{ a \tau_s}\right)^{2/3}.
	\label{eq:taubtilde}
\end{eqnarray}
These equations give the locus of the minimum as a perturbative expansion in $\alpha$.
 As a check on the validity of these computations we also perform a numerical study of the minima of the potential for different values of $\alpha$ and compare the results with our analytical computation. The results are plotted in Fig. \ref{fig:TAUbsPlots}.
	

We can see that for small values of $\alpha$ $(0<\alpha<3\times10^{-3})$  there is good agreement between the
 numerical location of the minimum coordinates and the approximate analytic solution.

Observation of Fig. \ref{fig:TAUbsPlots} reveals that the deviation between the numerical and the analytical grows with $\alpha$, but remains small (order of a few percent) throughout the range of values considered.
  The deviation between the analytical and the numerical results for $\tau_b$ is one order of magnitude larger than the one for $\tau_s$ - this
   is due to the exponential dependence of $\tau_b$ on $\tau_s$ as seen in Eq. (\ref{eq:taubtilde}), which amplifies the deviation.

\begin{figure}[ht]
\centering
\begin{minipage}[b]{0.45\linewidth}
\centering
\includegraphics[width=\textwidth]{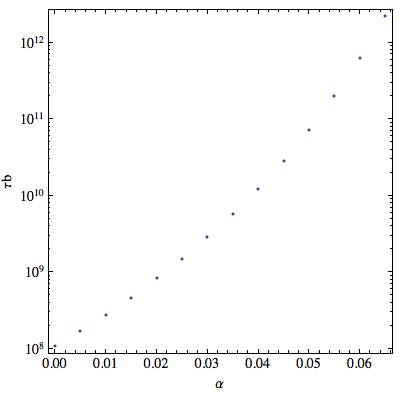}
\end{minipage}
\hspace{0.5cm}
\begin{minipage}[b]{0.45\linewidth}
\centering
\includegraphics[width=\textwidth]{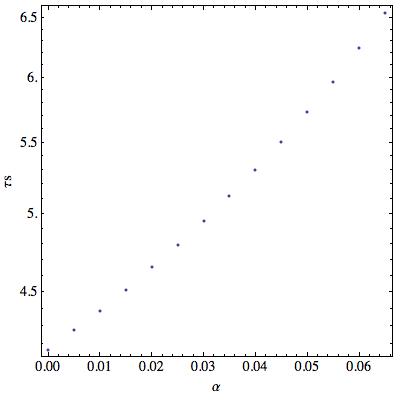}
\end{minipage}
\caption{$\tau_b (\alpha)$ and $\tau_s (\alpha)$: The linear expansion breaks down for higher values of $\alpha$.}
\label{fig:TAU12AnalyticPlot}
\end{figure}

As we consider larger values of $\alpha$ the linear approximation breaks down, as is illustrated in  Fig. \ref{fig:TAU12AnalyticPlot}. In this regime the effect of the $\alpha'^3$ also becomes negligible in the large volume limit.
However, we see that the structure of the potential (in particular the minimum at exponentially large values of the volume)
remains unaltered.
This is quite striking and shows that these two terms can play a similar role
in the stabilisation of the K\"ahler moduli of the theory.

\subsection{Redefining $\tau_b$}

We now consider the effects of redefining the large cycle $\tau_b$. In this case the motivation for the
redefinition is less strong
as neither of the arguments presented in section 3 directly apply to this cycle. However it is still worth considering for completeness.
We redefine the $\tau_b$ 4-cycle of $\mbb{P}^4_{[1,1,1,6,9]}$ according to Eq. (\ref{eq:redef}) and follow the same procedure
as in the previous section. The K\"ahler potential becomes:
\begin{equation}
	K=-2 \ln\left[\frac{1}{\lambda}([\tau_b-\beta \ln(\mathcal{V})]^{3/2}-\tau_s^{3/2})+\frac{\xi}{2}\left(\frac{S+\bar{S}}{2}\right)^{3/2}\right] + K_{cs}.
	\label{eq:Kredeftaub}
\end{equation}
The scalar potential is given by
\begin{equation}
	\begin{split}
	V=&\frac{3 |W_0|^2 \lambda^3 \xi}{4 g_s^{3/2} \tau_b^{9/2}}+\frac{9 |W_0|^2 \beta \lambda^2}{2 \tau_b^4}+\frac{8 a^2 |A|^{2} \lambda^2 \sqrt{\tau_s} e^{-2 a \tau_s}}{3 \tau_b^{3/2}}\\
	&-\frac{a \lambda^2 \tau_s e^{-2 a \tau_s}(6\beta \ln\mathcal{V}+3 \beta+2\tau_b)(A(\bar{A}+e^{a\tau_s} \bar{W_0})+c.c.)}{\tau_b^4},
	\end{split}
	\label{eq:}
\end{equation}
which can be further simplified to
\begin{equation}
	V=\frac{\Lambda}{\tau_b^{9/2}}+\frac{\Psi \beta}{\tau_b^{4}}+\Phi \frac{\sqrt{\tau_s} e^{-2 a \tau_s}}{\tau_b^{3/2}}-\theta \frac{\tau_s e^{-a \tau_s}}{\tau_b^3}\left(1+\frac{\beta}{\tau_b}(3 \ln\mathcal{V} +\frac{3}{2})\right),
	\label{eq:Vbeta}
\end{equation}
where
\begin{eqnarray}
	\Lambda &=& \frac{3 |W_0|^2 \lambda^3 \xi}{4 g_s^{3/2}},\\
	\label{eq:}
	\Psi &=&  \frac{9 |W_0|^2 \lambda^2}{2},\\
	\label{eq:}
	\Phi &=&\frac{8 a^2 |A|^2 \lambda^2}{3},\\
	\label{eq:}
	\theta &=& 2 a \lambda^2(A\bar{W_0}+ c.c.).
	\label{eq:}
\end{eqnarray}
In this case the effect of the redefinition in Eq. (\ref{eq:Vbeta}) is to create a term that dominates in the LARGE volume
limit over any other term in the potential (as it scales as $\sim \tau_b^{-4}$ while all other terms scale as $\sim \tau_b^{-9/2}$).
In the limit of small $\beta$ we find that the minimum of Eq. (\ref{eq:Vbeta}) is located at
\begin{equation}
	\tau_s=\tilde{\tau}_s\left(1+\frac{1}{3 a \tau_s}+\frac{\beta}{\tau_b}\left(\frac{-44}{9}\ln\mathcal{V}+\frac{16}{9}\frac{\tau_b^{3/2}}{\tilde{\tau}_s^{3/2}}\right)\right),
	\label{eq:}
\end{equation}

\begin{equation}
	\tau_b=\tilde{\tau}_b \left(1-\frac{1}{2 a \tau_s} +2\beta \frac{a \tau_s}{\tau_b}\right),
	\label{eq:}
\end{equation}
where $\tilde{\tau_s}$ and $\tilde{\tau_b}$ are given by Eqs. (\ref{eq:taustilde}) and (\ref{eq:taubtilde}) respectively.\\

\begin{figure}[ht]
\centering
\begin{minipage}[b]{0.45\linewidth}
\centering
\includegraphics[width=\textwidth]{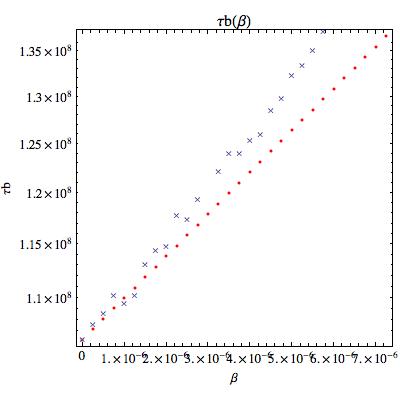}
\end{minipage}
\hspace{0.5cm}
\begin{minipage}[b]{0.45\linewidth}
\centering
\includegraphics[width=\textwidth]{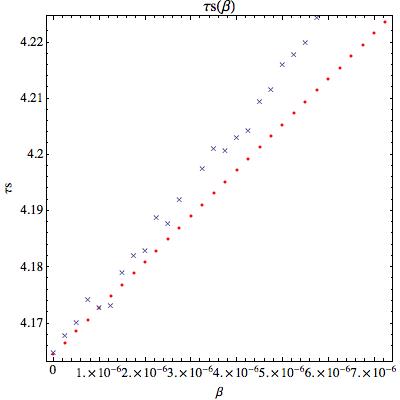}
\end{minipage}
\caption{$\tau_b$ and $\tau_s$ at the minimum as a function of $\beta$. The red dots are the analytical solution of Eqs.  while the blue dots represent the numerical result.}
\label{fig:TAUbsPlotsBeta}
\end{figure}

In Fig. \ref{fig:TAUbsPlotsBeta}  we depict the numeric and analytic location of the minima as functions of $\beta$. 
The conclusions are essentially the same as for the previously studied case, the main difference being the fact that the allowed range for the loop factor $\beta$ is much smaller.

In contrast to the case with $\tau_s$, in this case the LARGE volume minimum does not survive for large values of $\beta$. This is because
the correction to the scalar potential is dominant at large volumes over all other terms, and so gives a runaway behaviour. For sufficiently
large $\beta$ this runaway structure overwhelms all other features of the potential.

\subsection{Summary}

We have considered the effects of two possible field redefinitions on moduli stabilisation. In the better motivated case, that of the
redefinition of $\tau_s$, we find that the basic structure of the LARGE volume minimum is unaltered even though the effects of the redefinition
dominate at large volume the $\alpha'^3$ effects that played a crucial role in establishing the original LARGE volume minimum.
For completeness we have also studied the case of the redefinition of $\tau_b$. In this case the redefinition gives a term in the scalar potential
that at sufficiently large volume dominates all other terms and gives runaway.

\section{Soft SUSY Breaking}\label{sec:soft_supersymmetry_breaking_terms}

Having studied the effects of field redefinitions on moduli stabilisation let us now see whether they can give any new effects in the
study of supersymmetry breaking.

In the first formulation of the LARGE volume model it was originally assumed that the D7-brane supporting the MSSM fields and the D3-brane instanton wrap the same four-cycle, namely $\tau_s$. However it was pointed out in \cite{Blumenhagen:2007sm} that instanton zero mode counting forbids this scenario. It was proposed in \cite{Blumenhagen:2009gk} that the simplest way to avoid this issue is to assume a threefold with at least three four-cycles: one large cycle and two small del Pezzo four-cycles. The volume of this class of Calabi-Yau can be written as
\begin{equation}
	\mathcal{V}=(\eta_b \tau_b)^{3/2}-(\eta_s \tau_s)^{3/2}-(\eta_a \tau_a)^{3/2},
	\label{eq:Vol2}
\end{equation}
where $\tau_b$ determines the volume of the Calabi-Yau, $\tau_s$ is wrapped by the D3-brane instanton and the four-cycle of size $\tau_a$ supports the MSSM D7 brane. Examples of explicit Calabi-Yaus with these structures can be found in \cite{0404257, 08112936, 08114599}.
Since $\mbb{P}^4_{[1,1,1,6,9]}$ has only 2 K\"ahler moduli clearly it cannot be an example of a manifold obeying Eq. (\ref{eq:Vol2}). Nonetheless, the conclusions in the previous section regarding moduli stabilisation will remain the same once we consider a geometrical setup obeying Eq. (\ref{eq:Vol2}) as will be explained later.\\

 Let us start by reviewing the computation of soft terms following the review \cite{Brignole:1997dp}.
 This starts by performing an expansion of the K\"ahler potential and superpotential in powers of the matter fields :
\bea
	W & = & \hat{W}(\Phi)+\mu(\Phi) H_1 H_2+\frac{1}{6}Y_{\alpha\beta\gamma}(\Phi)C^{\alpha}C^\beta C^\gamma+...
	\label{eq:WExpand} \\
K & = & \hat{K}(\Phi,\bar{\Phi})+Z_{\alpha\bar{\beta}}(\Phi,\bar{\Phi}) C^\alpha C^{\bar{\beta}}+[Z(\Phi,\bar{\Phi})H_1H_2+h.c.]+...
	\label{eq:KExpand}
\eea
where $\Phi$ and $C^\alpha$ denote the moduli and the matter fields of the theory. In gravity mediation, the (scalar)
soft SUSY breaking terms are found by expanding the $\mathcal{N}=1$ scalar potential Eq. (\ref{eq:V}) in powers of the matter fields, using Eqs. (\ref{eq:WExpand}) and (\ref{eq:KExpand}).
The SUSY breaking is controlled by the F-terms of the moduli fields of the theory
\begin{equation}
	F^m=e^{\hat{K}/2}\hat{K}^{m\bar{n}}D_{\bar{n}}\bar{\hat{W}}.
	\label{eq:Fterm}
\end{equation}
From the fermionic part of the SUGRA Lagrangian we obtain the
canonically normalised gaugino field $\hat{\lambda}^{a} =\hbox{Re}(f_a)^{1/2}\lambda^a$
and gaugino masses given by
\begin{equation}
	M_{\tilde{G}}=\frac{1}{2 \hbox{Re}(f_i)}F^I\partial_I f_i.
	\label{eq:Mgaugino}
\end{equation}

The scalar Lagrangian can be written, for a diagonal matter metrics ($Z_{\alpha\bar{\beta}}=Z_{\alpha}\delta_{\alpha\bar{\beta}}$) as
\begin{equation}
	\mathcal{L}_{soft}=\tilde{K}_\alpha\partial_\mu C^\alpha \partial^\mu \bar{C}^{\bar{\alpha}}-M_{\tilde{Q}\alpha}^2 C^\alpha \bar{C}^{\bar{\alpha}}-\left(\frac{1}{6} A_{\alpha\beta\gamma}\hat{Y}_{\alpha\beta\gamma}C^{\alpha}C^{\beta}C^{\gamma}+B\hat{\mu}\hat{H}_1\hat{H}_2+h.c.\right),
	\label{eq:}
\end{equation}
where the scalar masses and A-terms are given by
\bea
	M_{\tilde{Q}}^2 & = & M_{3/2}+V_0-F^I F^{\bar{J}}\partial_I\partial_{\bar{J}}\ln Z_{\alpha}, \label{eq:MScalar} \\
	A_{\alpha \beta \gamma} & = & F^I \partial_I K + F^I \partial_IY_{\alpha\beta\gamma}-F^I\partial_I \ln(Z_\alpha Z_\beta Z_\gamma).
	\label{eq:ATerm}
\eea
Note that due to Peccei-Quinn shift-symmetry of the K\"ahler moduli, these fields are
forbidden to appear in the holomorphic superpotential. The superpotential Yukawa couplings,
$Y_{\alpha\beta\gamma}$, can therefore only depend on complex structure moduli.
Since flux stabilisation ensures that complex
structure moduli do not break SUSY (or at least the effects are highly suppressed),
the second term in Eq. (\ref{eq:ATerm}) vanishes.

The $\mu$ term is
\begin{equation}
	\hat{\mu}=\left(e^{K/2}+M_{3/2}Z-\bar{F}^{\bar{I}}\partial_{\bar{I}}Z\right)(Z_{H_1} Z_{H_2})^{-1/2}.
	\label{eq:muTerm}
\end{equation}

Finally the $B\mu$-term is given by
\begin{equation}
	\begin{split}
	B \hat{\mu}& =\frac{1}{\sqrt{Z_{H_1} Z_{H_2}}}\left(e^{K/2} \mu (F^I\partial_I K+F^I \partial_I \ln \mu-F^I \partial_I \ln(Z_{H_1}Z_{H_2})-M_{3/2})\right.\\
	 &+(2 M_{3/2}^2+V_ 0)Z - M_{3/2} \bar{F}^{\bar{I}}\partial_{\bar{I}}+M_{3/2}F^I(\partial_I Z -Z \partial_ I \ln(Z_{H_1}Z_{H_2})) \\
	&\left.-F^I F^{\bar{J}}(\partial_{\bar{I}}\partial_J Z - (\partial_{\bar{I}}Z)\partial_J \ln(Z_{H_1}Z_{H_2}))\right).
	\end{split}
\end{equation}

In order to compute these soft SUSY breaking terms one must specify the K\"ahler potential and superpotential for moduli fields $\hat{K}(\Phi,\bar{\Phi})$ and $\hat{W}(\Phi)$ in  Eqs. (\ref{eq:WExpand}) and (\ref{eq:KExpand}) as well as the gauge kinetic function $f_i$ and the matter metrics, $Z(\Phi)$. It is therefore crucial to note that there are two regimes in the effective field theory: the geometric regime and the singular cycle regime \cite{Blumenhagen:2009gk}. Our purpose here is to see whether moduli redefinitions can affects the results on supersymmetry
breaking that were found in \cite{Blumenhagen:2009gk}.\footnote{In \cite{Blumenhagen:2009gk} there is a cancellation of anomaly-induced soft 
terms that comes from the no-scale structure. This cancellation came from using conventional formulae for anomaly-mediated mass terms.
It has recently been argued in \cite{Shanta} that these formulae are not correct, and if this is the case then the cancellation used in
\cite{Blumenhagen:2009gk} will not apply. We note this issue but a detailed discussion of the form of anomaly-mediated soft terms is beyond
the scope of this paper.}

In the geometric regime all the four-cycles are larger than the string scale. In this regime the effective field theory is determined by:
\bea
	\hat{K}(\Phi,\bar{\Phi})& = & -2 \ln \left(\mathcal{V}+\frac{\hat{\xi}}{2}\right)-\ln(S+\bar{S})+K_{CS},
	\label{eq:GRKahler} \\
	\hat{W}(\Phi) & = & W_0+A e^{-a T_s}, \\
	\label{eq:GRW}
	f_{i} & = & T_a-\frac{1}{2}\kappa_i S.
	\label{eq:GRGKF}
\eea
where the volume is given by Eq. (\ref{eq:Vol2}) and the matter metric is
\begin{equation}
	Z=\frac{\tau_{a}^q f(\Phi)}{\tau_b^p},
	\label{eq:GRZ}
\end{equation}
with $f(\Phi)$ being a function of the complex structure moduli.
The form of the modular dependence of (\ref{eq:GRZ}) comes from taking the leading powers of the moduli.
Values of $p$ and $q$ can be either deduced or significantly constrained by
examining the behaviour of the physical Yukawa couplings under rescaling of the several moduli in the theory, as described in \cite{Conlon:2006tj}. Arguing that in local models the interactions should be independent of the overall volume it is found that $p_\alpha=1,  \forall \alpha$.
The $q$ is more delicate and can depend on whether a field originates from `internal' or `normal' modes of branes \cite{08052943}.
In what follows we leave $p$ and $q$ unspecified. This allows one to find more generic expressions for the soft terms and makes clear the nature of the cancellation of the leading order contributions.
	
In the singular cycle regime, the MSSM four-cycle is much smaller than the string scale. In this regime the theory is determined by:
\bea
	\hat{K}(\Phi,\bar{\Phi}) & = & -2 \ln\left(\mathcal{V}+\frac{\hat{\xi}}{2}\right)+c \frac{\tau_a^2}{\mathcal{V}}-\ln(S+\bar{S})
	\label{eq:SCRKahler} \\
	\hat{W}(\Phi) & = & W_0+A e^{-a T_s}
	\label{eq:SCRW} \\
	f_{i} & = & s_{ik} T_{k} +\delta_{i}S
	\label{eq:SCRGKF}
\eea
where the volume is given by
\begin{equation}
	\mathcal{V}=(\eta_b \tau_b)^{3/2}-(\eta_s \tau_s)^{3/2}
	\label{eq:Vol3}
\end{equation}
and the matter metric is
\begin{equation}
	Z=\frac{g(\Phi)+\tau_{a}^q h(\Phi)}{\tau_b^p},
	\label{eq:GRZ}
\end{equation}
with c being a constant and $g(\Phi)$ and $h(\Phi)$ being arbitrary functions of the complex structure moduli/dilation. The quadratic term in
$\tau_s$ in Eq. (\ref{eq:SCRKahler}) is introduced to ensure that  the $\tau_s$ kinetic term is well defined when $\tau_s = 0$.

\subsection{Redefinition of $T_a$ modulus and implications for soft SUSY breaking terms}

Our aim is to investigate the effects of the $T_a$ modulus redefinition in the scale of the soft SUSY breaking terms. In particular one seeks to compare the resulting terms to the ones found for the same geometrical configuration in \cite{Blumenhagen:2009gk}. One might also consider redefining the remaining K\"ahler moduli of the theory, but the effects will be subleading. We therefore neglect them and concentrate on the redefinition of $T_a$ only.\\

A fully consistent analysis must investigate the effects of the $T_a$ redefinition in the K\"ahler potential for moduli and in the matter metrics. We will proceed by steps, analysing first the case where we redefine $T_a$ in the K\"ahler potential only and then studying the full case.

\subsubsection{Redefining $T_a$ in the K\"ahler potential}\label{ssub:redefining_t_a_in_the_k}

First we investigate the consequences of applying the redefinition in Eq. (\ref{eq:redef}) to the K\"ahler potential for moduli fields, $\hat{K}(\Phi,\bar{\Phi})$. As the Standard Model cycle is taken to be stabilised by a D-term
we must impose the following condition for the K\"ahler modulus $T_a$:
\begin{equation}
	\partial_a K=0 \Leftrightarrow \tau_a=\alpha \ln \mathcal{V}.
	\label{eq:DTermCond}
\end{equation}
From this condition and since the superpotential is independent of $T_a$, it follows that $F_a=0$.
In \cite{Blumenhagen:2009gk} the vanishing of $F_a$ led to the vanishing of $F^a$.
However since the K\"ahler metric is non diagonal and $\tau_a \neq 0$ here we find
\begin{equation}
	\begin{split}
	F^a=&K^{a \bar{b}}F_{\bar{b}}+K^{a \bar{s}}F_{\bar{s}}\\
	&=-3\alpha M_{3/2}.
	\end{split}
	\label{eq:FA}
\end{equation}
This is a significant difference from \cite{Blumenhagen:2009gk}.
For the computation of the soft terms one also needs the F-terms for the remaining K\"ahler moduli. These are found to be
\bea
	F^b & = & -2 M_{3/2} \tau_ b,
	\label{eq:FB} \\
	F^s & = & -2 M_{3/2} \tau_ s,
	\label{eq:FS}
\eea
at leading order in the volume expansion. We note that even though the $\tau_a$ dependence of $\hat{K}(\Phi,\bar{\Phi})$ is different in the geometric and singular cycle regimes,  the results in Eqs. (\ref{eq:FA})-(\ref{eq:FS}) hold in both regimes.\\

In the computation of the soft SUSY breaking terms we will neglect the $\alpha'^{3}$ contributions to the K\"ahler potential as well as the nonperturbative superpotential. This is because the SUSY breaking structure of the LARGE volume models is essentially inherited from
no-scale, and the contributions of the $\alpha'$ corrections to the soft terms is volume suppressed
compared to the terms we will consider. \\

{\bf Geometric Regime}\\

Redefining the $\tau_a$ in the K\"ahler potential, Eq. (\ref{eq:GRKahler}), we find the soft terms to be given by:

\bea
	M_{\tilde{G}} & = & \frac{-3}{2}\alpha \frac{M_{3/2}}{\hbox{Re}(f_i)},
	\label{eq:GaginoMGR} \\
	M_{\tilde{Q}}^2 & = & M_{3/2}^2 (1-p)+\frac{9 q}{4} \frac{M_{3/2}^2}{\ln^2 \mathcal{V}}, \\
	\hat{\mu} & = & \frac{ Z}{\sqrt{Z_{H_1}Z_{H_2}}}M_{3/2}\left( (1-p+\frac{3}{2}\frac{q}{\ln \mathcal{V}})\right),
\eea
where we have set the superpotential $\mu$ term to vanish.	

The $B\hat{\mu}$ term is given in this regime by
\begin{equation}
	\begin{split}
	B\hat{\mu}=&\frac{M_{3/2}^2 Z}{\sqrt{Z_{H_1}Z_{H_2}}}\left((2-2p-p(p+1)+2p^2)+\frac{1}{\ln\mathcal{V}}3(1-p)(q_{H1}+q_{H2})+\right.\\
	&\left.\frac{1}{\ln^2\mathcal{V}}\frac{9 q}{2}(q_{H_1}+q_{H_2}-\frac{q-1}{2})\right)
	\end{split}
\end{equation}
where we have assumed that $\mu=0$ and that all the fields feel the overall volume of the CY in the same way, i.e., $p=p_{H_1}=p_{H_2}$.\\

The A-term in the geometric regime is given by
\begin{equation}
	A_{\alpha\beta\gamma}=M_{3/2}(3-3p)-3 M_{3/2}\frac{\tau_s^{3/2}}{\mathcal{V}}+\frac{3 M_{3/2}}{2 \ln \mathcal{V}}(q_\alpha+q_\beta+q_\gamma).
\end{equation}

Following the rescaling arguments in \cite{Conlon:2006tj}, we set $p=1$ and note that the leading order terms cancel in all soft terms, leaving only the volume suppressed contributions.\\

Neglecting $\mathcal{O}(1)$ factors we summarise the  results in the following table

\begin{center}
	\begin{tabular}{ c |c }

		Soft-term & Scale\\
		\hline
		\hline
		$M_{\tilde{G}}$& $M_{3/2}$\\
		\hline
		$M_{\tilde{Q}}^2$&$M_{3/2}^2/\ln^2\mathcal{V}$\\		
		\hline
		$\hat{\mu}$-term& $M_{3/2} \ln\mathcal{V}$\\
		\hline
		$B\hat{\mu}$-term&$M_{3/2}^2/\ln^2\mathcal{V}$\\		
		\hline
		A-term& $M_{3/2} \ln\mathcal{V}$\\		
		\hline
	\end{tabular}
\end{center}

{\bf Singular Cycle Regime}\\

We now compute the soft terms in the singular cycle regime. We start by defining
\begin{equation}
	\Gamma\equiv \frac{h (\Phi)\tau_a^q}{g(\Phi)+h(\Phi)\tau_a^q},
	\label{eq:Gamma}
\end{equation}
then the soft terms are:
\begin{equation}
	M_{\tilde{Q}}^2=M_{3/2}^2 (1-p)+\frac{M_{3/2}^2}{\ln^2\mathcal{V}}\frac{9 q}{4} \Gamma (q-1-q \Gamma),
		\label{eq:GaginoMSCR}
\end{equation}

\begin{equation}
	\hat{\mu}=\left(e^{K/2}\mu+M_{3/2} Z (1-p)+\frac{3 q}{2}\frac{M_{3/2}}{\ln \mathcal{V}}\Gamma\right)\frac{1}{\sqrt{Z_{H_1}Z_{H_2}}},
\end{equation}

\begin{equation}
\begin{split}
	B\hat{\mu}=\frac{M_{3/2}^2}{\sqrt{Z_{H_1}Z_{H_2}}}  \Big(&Z (2-2p-p(p+1)+p(p+1)) +\frac{1-p}{\ln\mathcal{V}}\left(\frac{3}{2} Z\Psi +3 pq \frac{\tau_{a}^{q}h(\Phi)}{\tau_{b}^{p}}\right) +\\
	&\frac{9 q/4}{\ln \mathcal{V}} \frac{\tau_{a}^{q}h(\Phi)}{\tau_{b}^{p}}(\Psi-q-1)\Big),
	\end{split}
	\end{equation}

\begin{equation}
	A_{\alpha\beta\gamma}=M_{3/2}(3-p_\alpha-p_\beta-p_\gamma)-3 M_{3/2}\frac{\tau_s^{3/2}}{\mathcal{V}}+\frac{3 M_{3/2}}{2 \ln \mathcal{V}} \sum_{\xi=\alpha,\beta,\gamma}q_\xi \Gamma_\xi,
\end{equation}
where
\begin{displaymath}
\Psi \equiv \frac{q_{1}h_{1}(\Phi)\tau_{a}^{q_{1}}}{g_{1}+h_{1}(\Phi)\tau_{a}^{q1}}+\frac{q_{2}h_{2}(\Phi)\tau_{a}^{q_{2}}}{g_{2}+h_{2}(\Phi)\tau_{a}^{q2}}.
\end{displaymath}

In the singular cycle regime the rescaling arguments used to find the modular dependence of the matter metric no longer apply. Nonetheless locality implies that $p=1$, then the leading order contributions to the soft terms vanish and the result found in this regime mirror the ones of the geometric regime up to factors of $\Gamma\approx \mathcal{O}(1)$.

The significance of this is that soft terms are of a similar order to the gravitino mass, in contrast to the behaviour in \cite{Blumenhagen:2009gk}
where soft terms were significantly suppressed compared to the gravitino mass.

\subsubsection{Redefining $T_a$ in the matter metrics}

In this section we study the effects on the scale of the soft terms of the MSSM four-cycle redefinition in the matter metrics. Once we take the redefinition of $\tau_a$ into account, the geometric regime matter metric becomes
\begin{equation}
	Z=\frac{(\tau_{a}-\alpha \ln\mathcal{V})^q f(\Phi)}{\tau_b^p},
	\label{eq:ZGR2}
\end{equation}
while in the singular cycle regime we should have
\begin{equation}
	Z=\frac{g(\Phi)+(\tau_a-\alpha \ln\mathcal{V})^q h(\Phi)}{\tau_b^p}.
	\label{eq:ZSCR2}
\end{equation}

It is crucial to note that, in general,  the metrics in Eqs. (\ref{eq:ZGR2}) and (\ref{eq:ZSCR2}) are singular and/or have singular derivatives once we impose the vanishing D-term condition. This poses a problem for the computation of soft SUSY breaking terms. A way to avoid this singular behaviour is to argue that the K\"ahler potential will receive higher order $\alpha'$ corrections which will modify the vanishing D-term condition. We sketch this idea in the geometric regime of the effective field theory and argue that the same argument holds in the singular cycle regime. Let $\phi(T_i)$ denote an arbitrary function of the K\"ahler moduli of the theory, then the full moduli K\"ahler potential for the $\mathcal{N}=1$ SUGRA in the geometric regime can be written as

\begin{equation}
		\hat{K}(\Phi,\bar{\Phi})=-2 \ln\left(\mathcal{V}+\frac{\hat{\xi}}{2}+\phi(T_i)\right)-\ln(S+\bar{S})+K_{CS},
\end{equation}
where we expect $\phi(T_i)\ll \mathcal{V}$. Then the vanishing D-term condition becomes
\begin{equation}
		\partial_a \hat{K}=0\Leftrightarrow \frac{\partial }{\partial \tau_a}\left(\mathcal{V}+\phi(T_i)\right)=0.
\end{equation}
After redefining $\tau_a$ in the volume this yields
\begin{equation}
	\sqrt{\tau_a-\alpha \ln\mathcal{V}}\left(1+\frac{3}{2}\frac{\alpha \sqrt{\tau_a}}{\mathcal{V}}\right)=-\frac{2}{3}\frac{\partial}{\partial \tau_a}\phi(T_i).
	\label{eq:Dterm}
\end{equation}
Since we do not know the explicit form of $\phi(T_i)$ we cannot solve Eq. (\ref{eq:Dterm}). Nonetheless one may write the solution as
\begin{equation}
	\tau_a=\alpha \ln \mathcal{V}+\theta,
\end{equation}
where $\theta \ll \alpha \ln\mathcal{V} $ is a function of the moduli of the theory, related to $\phi(T_i)$ by
\begin{equation}
	\sqrt{\theta}\approx-\frac{2}{3} \frac{\partial}{\partial \tau_a}\phi(T_i).
\end{equation}

Once these higher order corrections are taken into account, Eqs. (\ref{eq:ZGR2}) and (\ref{eq:ZSCR2}) no longer have a singular behaviour and the soft terms will be given as functions of $\theta$.\\

One must also point out that a full treatment of the problem does not involve a redefinition of $\tau_a$ in the gauge kinetic functions, Eqs. (\ref{eq:GRGKF}) and (\ref{eq:SCRGKF}), since these are protected by holomorphy. As a consequence of this, gaugino masses in the geometric and singular cycle regimes will still be given by Eqs. (\ref{eq:GaginoMGR})-(\ref{eq:GaginoMSCR}) respectively in the full case.\\

{\bf Geometric Regime}\\

Proceeding in the same way as before we compute the leading order contributions in the volume expansion of the soft terms:

\begin{equation}
	M_{\tilde{Q}}^2=M_{3/2}^2 \left(1-p-q_\alpha\frac{3 \alpha}{2 \theta}+\mathcal{O}(\mathcal{V}^{-1})\right),
	\label{eq:MScalarFull}
\end{equation}

\begin{equation}
	\hat{\mu}=\frac{M_{3/2} Z}{\sqrt{Z_{H_1}Z_{H_2}}}\left(1-p+\frac{3}{2}\frac{\alpha q_\alpha}{\theta}+\mathcal{O}(\mathcal{V}^{-1})\right),
\end{equation}

\begin{equation}
	B\hat{\mu}=\frac{Z}{\sqrt{Z_{H1}Z_{H2}}}M_{3/2}^2\left((p^2-3p+2)-\frac{3}{2}\frac{\alpha}{\theta}q_\alpha+\mathcal{O}(\mathcal{V}^{-1})\right),
\end{equation}

\begin{equation}
	A_{\alpha\beta\gamma}=3 M_{3/2}\left(1-p+\frac{1}{\mathcal{V}}\left(\theta^{3/2}+\frac{3}{4}\sqrt{\tau_a}\frac{\alpha^2}{\theta}(q_\alpha+q_\beta+q_\gamma)\right)+ \mathcal{O}(\mathcal{V}^{-2})\right).
\end{equation}

{\bf Singular Cycle Regime}\\

The soft terms in the singular cycle regime, once we redefine $\tau_a$ in the matter metric become:

\begin{equation}
	M_{\tilde{Q}}^2=M_{3/2}^2 \left(1-p-q_\alpha\frac{3 \alpha}{2 \theta} \tilde{\Gamma}+\mathcal{O}(\mathcal{V}^{-1})\right),
\end{equation}

\begin{equation}
	\hat{\mu}=\frac{M_{3/2} Z}{\sqrt{Z_{H_1}Z_{H_2}}}\left(1-p+\frac{3}{2}\frac{\alpha q_\alpha}{\theta}\tilde{\Gamma}+\mathcal{O}(\mathcal{V}^{-1})\right),
\end{equation}

\begin{equation}
	 B\hat{\mu}=\frac{Z}{\sqrt{Z_{H1}Z_{H2}}}M_{3/2}^2\left((p^2-3p+2)-\frac{3}{2}\frac{\alpha}{\theta}q_\alpha\tilde{\Gamma}+\mathcal{O}(\mathcal{V}^{-1})\right),
\end{equation}

\begin{equation}
	A_{\alpha\beta\gamma}=3 M_{3/2}\left(1-p+\frac{3}{4\mathcal{V}}\alpha^2\frac{\sqrt{\tau_a}}{\theta}\sum_{\xi=\alpha,\beta,\gamma}q_\xi\tilde{\Gamma}_{\xi}+\frac{c\theta^2}{2\mathcal{V}}+\mathcal{O}(\mathcal{V}^-1)\right),
	\label{eq:ATermFull}
\end{equation}
where $\tilde{\Gamma}_\alpha$ is defined as

\begin{equation}
	\tilde{\Gamma}_\alpha \equiv \frac{h (\Phi)\theta^q}{g(\Phi)+h(\Phi)\theta^q}.
\end{equation}

Even though an analysis with generic modular dependence in the matter metrics requires one to consider higher order $\alpha'$ corrections to the K\"ahler potential into account, there is a particular value of $q$ for which this is not strictly necessary. If one sets $q=2$ in the singular cycle regime, the matter metrics and its derivatives are well defined even if we set $\theta\rightarrow 0$. In this particular case we find, after setting $p=1$:

\begin{equation}
	\hat{\mu}=0=A_{\alpha\beta\gamma},
\end{equation}

\begin{equation}
	M_{\tilde{Q}}^2\propto \frac{\ln\mathcal{V}}{\mathcal{V}^3}\alpha^5,
\end{equation}

\begin{equation}
	B\hat{\mu}\propto \frac{\alpha^5 \ln \mathcal{V}}{\mathcal{V}^2}.
\end{equation}
In this case one also finds that the cancellation of the leading order terms happens as before, leaving only highly volume suppressed terms. The contribution to the soft terms computed in \cite{Blumenhagen:2009gk} will then dominate over the ones considered here.

Going back to the generic case, one must note that in both regimes, the usual cancelation of leading order terms is present if $p=1$. The scale of the soft terms is then parametrised by the ratio $\frac{\alpha}{\theta}$, where $\theta$ is essentially a derivative of the higher order $\alpha'$ terms in the K\"ahler potential for moduli fields, $\phi(T_i)$.

A full computation of the soft terms must include at least the terms in Eqs. (\ref{eq:MScalarFull})-(\ref{eq:ATermFull}) and the ones computed in \cite{Blumenhagen:2009gk}. To understand the relative size of both contributions and the scale of resulting the soft terms it is necessary to know $\phi(T_i)$ explicitly. At the moment this is beyond our possibilities. Nonetheless it is interesting to point out that it might be possible that the scale of the soft terms is set by the subleading terms in the K\"ahler potential. The main point here is to note that if we redefine the MSSM four-cycle according to Eq. (\ref{eq:redef}) the theory is still well behaved and we can get non vanishing soft terms.

\section{Conclusions}\label{sec:conclusions}

In this paper we have studied the effects of one-loop moduli redefinitions on moduli stabilisation in the
LARGE volume scenario. We have reviewed the origins of such redefinitions in orbifold models and also given a new
argument for the existence of such redefinitions in the geometric regime.
In our study of the effects on moduli stabilisation
we have assumed that the form of the K\"ahler potential is only altered by re-expressing the
geometric volume in terms of the redefined moduli variables. This leaves one obvious direction for future work
as it would be very interesting to find loop-corrected expressions for the K\"ahler potential including blow-up modes
(the results of \cite{0508043} apply only to the overall volume modulus).

We found that redefinitions of the small moduli do not alter the basic structure of the LARGE volume minimum: the
minimum remains in qualitatively the same location and at exponentially large volume. This is actually quite striking as
the redefinition generates terms that in the scalar potential at large volumes dominate the $\alpha'^3$ corrections.
For redefinitions of the overall volume, which is less motivated, the modified K\"ahler potential gives a scalar potential
that actually leads to runaway and removes the LARGE volume minimum.

We also studied the possible effects of moduli redefinitions on supersymmetry breaking. There we found that redefinitions
may have the ability to modify the results of  \cite{Blumenhagen:2009gk} and induce soft terms of a similar order to the
gravitino mass. However there were certain ambiguities which depend on the form of the matter metrics, and resolving these
ambiguities depends on corrections to the K\"ahler potential.
While the results of
this section may be potentially interesting, what is really needed is a full CFT computation to see the effect of redefinitions
on terms in the K\"ahler potential, which is however beyond the scope of this work.

\subsection*{Acknowledgments}

We thank Shanta de Alwis, Marcus Berg, Michael Haack, Eran Palti and Fernando Quevedo for useful discussions.
JC is supported by a Royal Society University Research Fellowship and by Balliol College, Oxford.
FGP is suported by Funda\c{c}\~{a}o para a Ci\^{e}ncia e a Tecnologia (Portugal) through the grant SFRH/BD/35756/2007.

\bibliographystyle{JHEP}

\end{document}